\documentclass[aps,showpacs]{revtex4}
\usepackage{graphicx}
\begin{document}
\title{Quantum $RLC$ circuits: charge discreteness and resonance}
\author{Constantino A. Utreras D\'{\i}az }
\affiliation{Instituto de F\'{\i}sica, 
Facultad de Ciencias, Universidad Austral de Chile, 
Campus Isla Teja s/n, Casilla 567, Valdivia, Chile}
\email{cutreras@uach.cl}
\date{\today, Valdivia} 
\begin{abstract}
In a recent article~\cite{UTRERAS}, we have advanced a semiclassical theory of quantum circuits with discrete charge and electrical resistance. In this work, we present a few elementary applications of this theory. For the zero resistance, inductive circuit, we obtain the Stark ladder energies in yet another way; and generalize earlier results by Chand\'{\i}a et. al~\cite{CHANDIA}, for the circuit driven by a combination d.c. plus a.c. electromotive force (emf).  As a second application, we investigate the effect of electrical resistance, together with charge discreteness, in the current amplitude, and resonance conditions of a general $RLC$ quantum circuit, including nonlinear effects up to third order on the external sinusoidal emf.
\end{abstract}
\maketitle
\section{Introduction}

In a series of articles Li and Chen~\cite{LI-CHEN,YOU-LI}  and us~\cite{FLORES,FLORES-UTRERAS,FLORES-UTRERAS2,UTRERAS-FLORES,FLORES-BOLOGNA}, have developed a theory of quantum electrical systems, based on a treating such systems as quantum $LC$ circuits; that is, electrical systems described by two fenomenological parameters: an inductance $L$, and a capacitance $C$. Such quantum theory of circuits is expected to apply when the transport dimension becomes comparable with the charge carrier coherence length, taking into account both the quantum mechanical properties of the electron system, and also the discrete nature of electric charge. Now, in a recent work~\cite{UTRERAS}, we have proposed a {\em semiclassical} theory of quantum electrical circuits. This has been done with two goals in mind. First, to obtain useful predictions of the theory from very simple calculations, obtaining for example energy spectra, and, second, to push the circuit analogy one step further,generalizing the equations of motion to include electrical resistance. As it is well known, the problem of electrical resistance, at the mesoscopic level, is that of the contact resistance, i.e., the coupling between the mesoscopic system and a macroscopic system with which it interacts, which leads to the conductance quantization condition, as it was first shown by Landauer~\cite{IMRY}.

The {\em semiclassical} theory of quantum $RLC$ circuits~\cite{UTRERAS} starts from the quantum Hamiltonian of the $LC$ circuit~\cite{UTRERAS,LI-CHEN,YOU-LI,FLORES,FLORES-UTRERAS,FLORES-UTRERAS2,UTRERAS-FLORES,FLORES-BOLOGNA}, and adds a term to account for electrical resistance, $R \phi/L$. The resulting equations become
\begin{eqnarray}
\label{eqRes2}
-\dot \phi - R \frac{\phi}{L}   &=& \frac{q}{C} \\
\dot q & =& \frac{\phi_0}{L} \sin(\phi/\phi_0). \nonumber
\end{eqnarray}
The equations above are considered, mathematically, as classical equations, but they include quantum effects, the quantized nature of electric charge (through the parameter $\phi_0 = \hbar/q_e$), and electrical resistance $R$; which couples the system to the external reservoir~\cite{UTRERAS}. These equations are highly nonlinear, but they reduce to the usual equations of the $RLC$ circuit, in the discrete charge limit, $q_e \to 0$. An interesting consequence is that it is possible to find stable, constant charge and flux ($ \phi = 2 n \pi \phi_0$) solutions of the equations~\ref{eqRes2}, such that, if one further assumes that the charge is discrete ($q = -m q_e$), then the following relation should exist between the system parameters~\cite{UTRERAS}, which bears a strong resemblance to the Landauer formula~\cite{IMRY}

\begin{eqnarray}
G_{eff} = \frac{R C }{L} = \frac{m}{n} \frac{q_e^2}{h} = \frac{m}{n} G_{Landauer}.
\end{eqnarray}
In this work, we apply our theory to some simple systems; in section~\ref{secStark}, we consider the so-called $L$ -design model, a quantum circuit under a constant electromotive force $\varepsilon$, obtaining the Stark ladders in a very simple manner~\cite{CHANDIA}. Next, in section~\ref{secRL}, we consider the same $L$ design case, add electrical resistance, according to our semiclassical theory, and subject it to a combined a.c. plus d.c. electromotive force, $\varepsilon (t) = \varepsilon_0 + \varepsilon_1 \cos (\omega t)$. 
In section~\ref{secRLC}, we consider a quantum $RLC$ circuit under a sinusoidal-only external electromotive force $\varepsilon(t) = \varepsilon_0 \cos(\omega t)$, and study the resonance conditions, considering nonlinear effects up to $\varepsilon_0^3$, and electrical resistance. We find that the resonance frequency of the quantum dissipative system is lowered, with respect to the non dissipative system, the shift is found to be $\Delta \omega /\omega_0 = - (\varepsilon_0/R)^2/(4\phi_0/L)^2$. We remark that the results obtained show the usefulness of the semiclassical approach to both dissipative and non-dissipative quantum circuits, therefore, it provides a way to advance our understanding of quantum circuits, and encourages us to pursue the subject in future works.

\section{Inductive circuit: stark ladders}
\label{secStark}

In a recent work, Chand\'{\i}a et. al.~\cite{CHANDIA}, using the quantum circuit approach have obtained the energies of the Stark ladders, previously conjectured by Chen at. al.~\cite{CHEN-SHEN}. Chand\'{\i}a et. al.~\cite{CHANDIA} consider a quantum circuit without resistance, in the $L$ design situation (the capacity, $C \to \infty$) situation, under a constant applied electromotive force $\varepsilon$. In this work, we that the semiclassical method described in a previous article ~\cite{UTRERAS} may be used to obtain the quantized energies of the stark ladders. To prove this, consider a quantum circuit (and large, $C\to \infty$ capacity), under a d.c. electromotive force $\varepsilon$. Using $\phi_0 = \hbar/q_e$, the equations of motion may be written as

\begin{eqnarray}
\varepsilon - \frac{d \phi}{dt} & = & 0 \\
\label{Eq:carga}
\frac{dq}{dt} & = & \frac{ \phi_0}{L} \sin (\phi/\phi_0),
\end{eqnarray}
and the Hamiltonian is
\begin{equation}
H = - q \varepsilon + \frac{2 \phi_0^2}{L} \sin^2(\phi/2\phi_0).
\end{equation}
To find the energy spectrum, we use the semiclassical method~\cite{UTRERAS}, finding first the charge $q$ as a function of the energy, 
\begin{equation}
q = \frac{1}{\varepsilon} \left( -E  + \frac{2 \phi_0^2}{L} \sin^2(\phi/2\phi_0) \right).
\end{equation}
Next, we compute the action variable $J$,
\begin{equation}
J = \oint q d\phi = \int_0^{2\pi \phi_0} q d\phi = -\frac{2\pi \phi_0}{\varepsilon} \left( E - \frac{\phi_0^2}{L} \right),
\end{equation}
and impose the quantization condition in the semiclassical way (Bohr-Sommerfeld), $J = n h$, we obtain precisely the result of Chen et al ~\cite{CHEN-SHEN}, also obtained by Chandia et al.~\cite{CHANDIA}, namely
\begin{equation}
E = - n q_e \varepsilon +  \frac{\phi_0^2}{L}.
\end{equation}
It it interesting to see that these results are very robust, in the sense that they are preserved, under very different mathematical descriptions, depending strongly on the charge quantization condition more that anything else.

\section{Inductive circuit: combined a.c., d.c. and resistance}
\label{secRL}

In their recent article, Chandia et. al~\cite{CHANDIA} consider a quantum inductive, nondisipative circuit, subject to a  weak external electromotive force given by   $\varepsilon (t) = \varepsilon_0 + \varepsilon_1 \cos(\omega t )$. They showed that the (physical) electrical current in such a quantum circuit, in the first order approximation, has a nonzero time average over a period $T = 2\pi/\omega_B$, given by

\begin{equation}
\label{Eq:average}
< \frac{dq }{dt} > = \frac{\varepsilon_1 \omega_B}{\pi L} \left(  \frac{\sin^2 (\pi \omega/\omega_B     ) }{\omega_B^2 - \omega^2}            \right).
\end{equation}
In this equation $\omega_B = q_e \varepsilon_0/\hbar = \varepsilon_0/\phi_0$ ($\phi_0 = \hbar/q_e$ is a flux quantum). To arrive to equation~(\ref{Eq:average}), one computes the flux, replaces into equation~(\ref{Eq:carga}), and expands the sine function up to first order, assuming that the alternating voltage $\varepsilon_1$ is small, i.e. $\varepsilon_1 << \omega \phi_0 $,

\begin{equation}
\frac{dq}{dt} =  \frac{\phi_0}{L} \left(  \sin ( \omega_B t )  + 
 \frac{ \varepsilon_1 \omega_B}{\varepsilon_0 \omega } \sin (\omega t) \cos (\omega_B t  )  \right) .
\end{equation}
Equation~(\ref{Eq:average}) is obtained after averaging equation (\ref{Eq:average})over a time period $T = 2\pi/\omega_B$. The results ~\cite{CHANDIA}, tell us that
\begin{itemize}
\item The electric current has zero average when $\omega = n \omega_B$, for integer $n$.
\item The electric current has nonzero average for $\omega \ne n \omega_B$, and it shows extrema (resonances) when $\omega \approx (n + 1/2) \omega_B$; the strongest maximum occurs for $\omega \approx \omega_B/2$.
\item For  small frequencies, $\omega << \omega_B$, 
\begin{eqnarray}
< \dot q > &=& \frac{\pi \varepsilon_0   \omega^2 }{L \omega_B^3}.
\end{eqnarray}
\end{itemize}

Another important observation is that, in alternating current circuits, it is customary to average with respect to the pariod of the source, and also that the observable quantity would be the root mean square of the current, $ \sqrt{ < \dot{ q}^2 > } $, instead of the average value of the current. First, define the parameter $ k = \varepsilon_1 \omega_B /(\varepsilon_0\, \omega )$, then compute the average, for the square of the current, given below, neglecting $k^2$ terms,

$$
\dot q^2 = ( \frac{\phi_0}{L} )^2 \left( \sin^2(\omega_B t) + k \sin(2 \omega_B t) \sin(\omega t)  \right).
$$
The results are,

\begin{itemize}
\item Average for $T=2 \pi/\omega_B$. 

\begin{equation}
< \dot q^2> = \frac{\phi_0^2}{2 L^2} \left( 1 + \frac{ 2 k \omega_B^2 \sin(2\pi \omega/\omega_B) }{\pi (4\omega_B^2 - \omega^2) }  \right)
\end{equation}
\item Average for $T = 2\pi/\omega$

\begin{equation}
< \dot q^2>  = \frac{\phi_0^2}{2 L^2} \left(  1 - \frac{ \omega \sin(4 \pi \omega_B/\omega ) }{4 \pi \omega_B}  + 
\frac{ k \omega^2 \sin( 4 \pi \omega_B/\omega) }{\pi (4\omega_B^2 - \omega^2) }    \right)
\end{equation}

\end{itemize}

Now, let us study the changes that may be observed when one includes electrical resistance in our description of the inductive, discrete-charge, quantum circuit discussed previously. As it has been shown by previously~\cite{UTRERAS}, the equations that describe an inductive circuit {\em with electrical resistance} under the emf $\varepsilon_0 + \varepsilon_1 \cos(\omega t )$ are

\begin{eqnarray}
\dot \phi + \frac{R}{L} \phi &=& \varepsilon_0 + \varepsilon_1 \cos(\omega t ) \\
\dot q & = & \frac{\phi_0}{L} \sin (\frac{\phi}{\phi_0} ).
\end{eqnarray}
Since these equations remain uncoupled, they may reduced to an integration. The particular solution  for $\phi = \phi_p(t)$ may be written as 

\begin{equation}
\label{Eq:part}
\phi_p(t) = A + B \cos(\omega t - \alpha).
\end{equation}
Use the {\em complex impedance} $ Z(\omega) = R + i \omega L$, and $|Z(\omega)| = \sqrt{ R^2 + ( \omega L)^2 }$, and the phase angle $\alpha$, $\tan (\alpha ) = \omega L /R$, then  the coefficients $A$ and $B$ become

\begin{eqnarray}
A & =& \frac{L\varepsilon_0 }{R} \\
B & = & \frac{L \varepsilon_1}{ |Z(\omega)| }
\end{eqnarray}
The solution obtained so far does not take into account the initial condition at $t=0$, therefore we must add an exponentially decaying term, $e^{-t/\tau}$, $\tau = L/R$, so that

\begin{equation}
\phi(t) = A + B \cos(\omega t - \alpha) + C e^{-t/\tau}.
\end{equation}
The constant $C$ is chosen so that $\phi (t=0) = 0$, then $C = - A - B \cos(\alpha)$, then the proper solution is

\begin{equation}
\phi(t) = A \, \left( 1 - e^{-t/\tau} \right)   + B \, \left(  \cos(\omega t - \alpha) - \cos(\alpha) e^{-t/\tau} \right) .
\end{equation}
This solution has the correct behaviour as $R\to 0$, namely, it coincides with the solution  from the $R=0$ case, equation (\ref{Eq:carga}); however, there are significant  differences, since:
\begin{itemize}
\item The solution  $\phi(t)$ for the case $R=0$ increases linearly with time, while, for $R\ne 0$, $\phi(t)$ grows linearly with time {\em only for $t << \tau $}
\item For $t >> \tau$, and $R\ne 0$, the solution $\phi (t )$ is always bound, and it coincides with $\phi_p(t)$, in equation (\ref{Eq:part}).
\end{itemize}
We conclude that the effect described by Chandia et. al~\cite{CHANDIA} may only be observed for very weakly dissipative systems, for short times, $t << \tau $, so that the observation times involved ($2\pi/\omega$ and $2\pi/\omega_B$) should be much shorter than the characteristic time $\tau = L/R$. For longer times, $t>>\tau$, the current will be

\begin{equation}
\dot q  = \frac{\phi_0}{L} \sin \left(   \frac{L\varepsilon_0 }{R \phi_0 }  +    
\frac{L \varepsilon_1}{ |Z(\omega)| \phi_0 }  \cos(\omega t - \alpha)    \right).
\end{equation}
Now, let us consider the case in which $L \varepsilon_1/|Z(\omega )| \phi_0 << 1$, then 

\begin{equation}
\label{Eq:qrl}
\dot q  = \frac{\phi_0}{L} \left[  \sin \left(   \frac{L\varepsilon_0 }{R \phi_0 } \right)  +
\frac{L \varepsilon_1}{ |Z(\omega) | \phi_0 } \cos \left(   \frac{L\varepsilon_0 }{R \phi_0 } \right) 
 \cos(\omega t - \alpha)   \right] ,
\end{equation}
so that the time average of the electric current, over a period of the source, $ T = 2\pi/\omega$, is
\begin{equation}
< \frac{dq}{dt}>  = \frac{\phi_0}{L} \sin \left(   \frac{L\varepsilon_0 }{R \phi_0 } \right).
\end{equation}
Notice that the current $\dot q$ has nonzero time average, except for some particular values of the argument of the sine function,
$L \varepsilon_0 /R\phi_0 = n \pi$, in other words, $\varepsilon_0 = n \pi \phi_0/\tau$, a very simple result.

Now, for $t>>\tau$, the solution for $\dot q$ is given by equation (\ref{Eq:qrl}), therefore, averaging over a period of the source ($T = 2\pi/\omega$)

\begin{equation}
<\dot q^2> = \frac{\phi_0^2}{L^2} \left( \sin^2 \left( \frac{L \varepsilon_0}{R \phi_0} \right) +  \frac12 \left( \frac{L \varepsilon_1}{|Z(\omega)| \phi_0}  \right)^2  
\cos^2 \left( \frac{L \varepsilon_0}{R \phi_0} \right) \right).
\end{equation}

\section{RLC circuit: resonance}
\label{secRLC}

Consider now the circuit under a purely sinusoidal perturbation, $\varepsilon (t) = \varepsilon_0 \cos(\omega t)$, and now look for a solution as a pertubative series, we obtain  only the first two terms of the series. We write down the circuit equations once again, 

\begin{eqnarray*}
\dot \phi + \frac{R}{L} \phi + \frac{q}{C} & = & \varepsilon_0 \cos(\omega t) \\
\dot q &=& \frac{\phi_0}{L} \sin \left( \frac{\phi}{\phi_0} \right).
\end{eqnarray*}
Let us seek solutions as 

\begin{eqnarray*}
q (t) &=& q_1(t) + q_3(t) + \cdots \\
\phi(t) &=& \phi_1(t) + \phi_3(t) + \cdots ,
\end{eqnarray*}
we write them like this since we know that upon series expansion the solution $q_1$ (and $\phi_1$) will be linear on the parameter $\varepsilon_0$, and the solution $q_3$ (and $\phi_3$) will be proportional to $\varepsilon_0^3$, with only odd-order terms appearing. Now, insert the series  solutions into the $RLC$ equations, obtaining the linearized equations

\begin{eqnarray}
\dot \phi_1 + \frac{R}{L} \phi_1 + \frac{q_1}{C} & = & \varepsilon_0 \cos(\omega t ) \\
\dot q_1 & = & \frac{\phi_1}{L} \\
\dot \phi_3 + \frac{R}{L} \phi_3 + \frac{q_3}{C} & = & 0 \\
\dot q_3 & = & \frac{\phi_3}{L} - \frac{\phi_1^3}{6 \phi_0^2 L }.
\end{eqnarray}

It is easy to see that both equations may be solved by the complex-number method. Therefore it is convenient to define the complex impedance $Z(\omega)= R + i\omega L + 1/i\omega C = |Z(\omega)| e^{i\alpha_1}$, as well as its phase angle and magnitude,
\begin{eqnarray}
Z (\omega ) & = & R + i\omega L + 1/i\omega C \\
|Z(\omega)| & = & \sqrt{ R^2 + (\omega L - 1/\omega C)^2 } \\
\tan (\alpha_1 ) & = & \frac{\omega L - 1/\omega C}{R}.
\end{eqnarray}
In this way, the first order pseudo flux $\phi_1$ and pseudo current are

\begin{equation}
I_1(t) = \frac{\phi_1 (t) }{L} =  \frac{\varepsilon_0 }{|Z(\omega)|} \cos(\omega t - \alpha_1).
\end{equation}
To write the current $\dot q$, we define the  parameters,
\begin{eqnarray}
m & = & \frac{(\phi_1^0 )^3}{6 \phi_0^2 L } =  \frac{L^2}{6 \phi_0^2}  \frac{\varepsilon_0^3}{|Z(\omega)|^3 } \\
\tan (\alpha_3 ) &= & \frac{3\omega L - 1/3\omega C}{R}\\
\tan (\beta_1 ) &= & \frac{\omega L}{R} \\
\tan (\beta_3) & = & \frac{3\omega L}{R} 
\end{eqnarray}
The current $\dot q$ below contains terms oscilating at the frequency $\omega$, and also at $3\omega$, which comes from the cubic nonlinearity, 
\begin{eqnarray*}
\dot { q } (t) & = &\frac{\varepsilon_0}{ |Z(\omega)| } \cos ( \omega t - \alpha_1) - \frac{3 m \sqrt{ R^2 + ( \omega L)^2 }  }{ 4 |Z(\omega)| } \cos (\omega t - 2 \alpha_1 + \beta_1 )  + \\
& &  - \frac{m \sqrt{ R^2 + (3\omega L)^2} }{4 |Z(3\omega)| } \cos ( 3 \omega t - 3 \alpha_1 - \alpha_3 + \beta_3) .
\end{eqnarray*}

The results of the previous paragraphs allow us to see that there is a slight change of the resonace frequency for a quantum circuit, the corrections are small, and depend in a nonlinear way on the external source amplitude $ \varepsilon_0$. Define the dimensionless parameter $\kappa <<1$,

\begin{equation}
\kappa  =  \frac{3 m \sqrt{ R^2 + ( \omega L)^2 }  }{ 4 \varepsilon_0 } = 
\frac{ \sqrt{ R^2 + ( \omega L)^2 } (L\varepsilon_0)^2 }{8 \phi_0^2 |Z(\omega)|^3},
\end{equation}
the terms oscilating at frequency $\omega$ may be put together in a single term, then the current may be written as (defining the additional phase angle $\Delta \theta$). We neglect the term at frequency $3\omega$, since if one evaluates its root mean square average for a period of time $T = 2\pi/\omega$, it will turn out a zero value.

\begin{eqnarray}
\dot q(t) &=& \frac{\varepsilon_0}{ |Z(\omega)| }  
\sqrt{  1  + \kappa^2 - 2  \kappa \cos (- \alpha_1 + \beta_1 ) }  \cos  (\omega t - \alpha_1  + \Delta \theta). \\
& \approx & \frac{\varepsilon_0}{ |Z(\omega)| }  
\left[  1  -  \kappa \cos (- \alpha_1 + \beta_1 ) \right]   \cos  (\omega t - \alpha_1  + \Delta \theta) \\
& = & I_0 (\omega)  [  1  -  A (\omega) ]   \cos  (\omega t - \alpha_1  + \Delta \theta),
\end{eqnarray}
where we have defined the constants

\begin{eqnarray}
I_0(\omega) & = & \frac{\varepsilon_0}{|Z(\omega)|}  \\
A(\omega) & = & \kappa \cos (\beta_1 - \alpha_1) \\
I(\omega) & = & I_0 ( 1 - A(\omega)).
\end{eqnarray}
We  may put out main result, $I(\omega)$, in simpler terms, to do that, use the addition formula for cosine, and the definitions given earlier for the paramenters $\beta_1$, $\alpha_1$, etc.; after some algebra, we get 

\begin{eqnarray}
\cos (\beta_1 - \alpha_1) &=& \frac{R^2 + \omega L (\omega L - 1/\omega C)}{\sqrt{R^2 + (\omega L)^2} |Z(\omega)|} \\
A(\omega ) &=& \frac{ (L \varepsilon_0)^2 }{8 \phi_0^2 |Z(\omega)|^4}  \left[  R^2 + \omega L (\omega L - 1/\omega C)   \right].
\end{eqnarray}

From this we shall compute the shift on the resonance frequency due to both charge quantization and electrical resistance. To do this, we expand the equation above about $\omega = \omega_0 =1/\sqrt{L C}$, up to second order,

\newcommand{\dw}{(\omega - \omega_0)}
\begin{eqnarray}
A(\omega)  &=& \frac{ (L \varepsilon_0)^2 }{8 \phi_0^2 R^2} 
\left[  1 +  \frac{2 L \dw }{\omega_0 C R^2} - \frac{7 L^2 \dw^2}{R^2} + \cdots   \right] \\
I_0(\omega) & = & \frac{\varepsilon_0}{R} \left[ 1 - \frac{2 L^2 \dw^2}{R^2} + \cdots \right].
\end{eqnarray}
\newcommand{\Dw}{\Delta \omega}
We need to solve $dI/d\omega = 0$, for $\Dw = \omega - \omega_0$, recalling that $A(\omega) <<1$, and that $dI_0(\omega_0)/d\omega = 0$,

\begin{eqnarray}
\frac{dI (\omega) }{d\omega} & = & \frac{dI_0 (\omega) }{d\omega} ( 1 - A(\omega) ) - 
I_0(\omega) \frac{dA(\omega) }{d\omega} \\
\frac{dI (\omega) }{d\omega} & = & \frac{d^2 I_0 (\omega_0) }{d\omega^2 } \Dw  ( 1 - A(\omega_0) ) - 
I_0(\omega_0) \left( \frac{dA(\omega_0) }{d\omega} + \frac{d^2 A(\omega_0) }{d\omega^2} \Dw \right),
\end{eqnarray}
using the expansions above, we obtain the solution
\begin{equation}
\Delta \omega = - \frac{\omega_0 (L \varepsilon_0)^2}{16 \phi_0^2 R^2} = -\frac{ (\varepsilon_0/R )^2 \, \omega_0  }{ 16 (\phi_0/L)^2 }  
\end{equation}

\section{Final Remarks}

In this article, we have studied some simple applications of the ideas of discrete-charge quantum $RLC$ circuits, and applied to two simple systems. In the first place, we reobtained the energy spectrum (Stark ladders) for the model $L$-design system, using the semiclassical method~\cite{UTRERAS}. Next, we consider the $L$ design case studied by Chand\'{\i}a~\cite{CHANDIA}, but including the effect of external resistance, finding that their results hold under more stringent conditions, namely that the observation times should be much shorter that the characteristic time $\tau = L/R$; finding the correction in the other case. Finally, we used the generalized quantum $RLC$ circuit equation to study the $RLC$ circuit under a sinosoidal electromotive force. We have shown that the resonance frequency of the system becomes shifted, and the shift is proportional to the square of the ratio of a {\em classical current} $\varepsilon_0/R$ and a {\em quamtum current} $\phi_0/L$; a simple result may be tested under appropiate experimental conditions.

\section{Acknowledgements}

The author aknowledges the finantial support provided by DIDUACH Grant \# S-2004-43. Thanks are due to Prof. J. C. Flores, for discussions, comments, and the kind hospitality of the Instituto de Alta Investigaci\'on of the Universidad de Tarapac\'a.

\end{document}